\documentclass[]{JHEP3} 

\newcommand{\be}{\begin{equation}}
\newcommand{\ee}{\end{equation}}
\newcommand{\bea}{\begin{eqnarray}}
\newcommand{\eea}{\end{eqnarray}}
\newcommand{\bvec}[1]{\mbox{\boldmath $#1$}}
\newcommand{\eqref}[1]{(\ref{#1})}

\title{Towards the String representation of the dual
Abelian Higgs model  beyond the London limit}
\author{Yoshiaki Koma\\
	Institute for Theoretical Physics,  Kanazawa University, \\
	Kanazawa 920-1192,  Japan.\\
	E-mail: \email{koma@hep.s.kanazawa-u.ac.jp}
}
\author{Miho Koma(Takayama), Dietmar 
Ebert\thanks{On leave of absence from Institute of Physics, Humboldt 
University, Berlin} and 
Hiroshi Toki\\
	Research Center for Nuclear Physics, Osaka University, \\
	Mihogaoka 10-1, Ibaraki, Osaka 567-0047, Japan \\
	E-mail: \email{takayama@rcnp.osaka-u.ac.jp} \\
	E-mail: \email{ebert@rcnp.osaka-u.ac.jp} \\
	E-mail: \email{toki@rcnp.osaka-u.ac.jp}
}

\preprint{}

\abstract{
  We perform a path-integral analysis of
  the string representation of the dual Abelian Higgs (DAH)
  model beyond the London limit, where the string describing the
  vortex of a flux tube has a finite thickness. We show that besides
  an additional vortex core contribution to the string tension, 
  a modified Yukawa interaction appears as a 
  boundary contribution in the type-II dual
  superconducting vacuum. In the London limit, the modified Yukawa
  interaction  is reduced to the Yukawa one.
}

\begin{document}

\setlength{\baselineskip}{0.49cm}

\section{Introduction}\label{sec:Intro}

\par
The construction of a realistic low-energy Lagrangian for hadrons
based on quantum chromodynamics (QCD) clearly requires a deeper
understanding of the mechanism of confinement.  A very useful concept
for an analytical description of this phenomenon is the dynamical
scheme of a dual superconductor proposed more than twenty years ago by
't~Hooft and Mandelstam \cite{tHooft-1976,mandelstam}.  This approach
emphasizes, in particular, the role of magnetic monopoles for
confinement.  The condensation of monopoles squeezes the
chromoelectric flux into (open) Abrikosov-Nielsen-Olesen (ANO) type
vortices \cite{abrikosov-1957,nielsen-1973,nambu-1974},
which then confine the quark and antiquark
sitting at their ends.  Recent studies in lattice QCD in the maximally
Abelian gauge indeed suggest remarkable properties of the QCD vacuum,
such as Abelian dominance \cite{tsuzuki-1990,brandstater-1991}
and monopole condensation
\cite{kronfeld-1987}, which 
confirm the dual superconductor picture, suitably
described by a dual Abelian Higgs (DAH) model representing just a
corresponding dual Ginzburg-Landau type of theory \cite{tsuzuki-1988,
suganuma-1995-npb}. 
The DAH model is here obtained by Abelian projection \cite{tHooft-1981}
which is a crucial step in order to find the relevant IR degrees of
freedom of QCD.

\par
In the present paper we are mainly interested in analytical studies of
the confinement mechanism leading to a path-integral derivation of
effective string  actions from the $SU(2_c)$-DAH model. The corresponding
path-integral approach is essentially simplified in the so-called
London limit of large monopole self-coupling $\lambda \to \infty$ (large
monopole mass $m_{\chi} \to \infty$), where the modulus of the
magnetic monopole field is frozen to its v.e.v., and ANO vortices become
infinitely thin (core radius $ \rho
= (m_{\chi})^{-1}\to 0$). In this limit the DAH
model can be suitably reformulated as a theory of a massive
antisymmetric Kalb-Ramond (KR) field interacting with surface elements
of the world-sheet swept out by the ANO vortex (Dirac string).
By performing finally a derivative expansion of the resulting
effective action, it was possible to derive the Nambu-Goto action
\cite{lee-1993,orland-1994,sato-1995,akhmedov-1995}
 including a correction (rigidity) term 
 \cite{polyakov-1986,kleinert-1986},
 to estimate field strength correlators of the DAH
model \cite{antonov-1998,
baker-1998} and comparing them with corresponding
quantities of the Stochastic Vacuum Model of QCD \cite{dosch-1987,
simonov-1987,antonov-1996}.
Obviously, the London limit picture is only appropriate for large transverse
distances from the vortex (string), where the  thickness of the core and the
corresponding contribution  to the field energy per
unit length (string tension) are neglected \cite{lee-1993,orland-1994,
sato-1995,akhmedov-1995,
antonov-1998}.
Moreover, since the monopole
field in this approximation is nowhere vanishing, one gets a massive
dual gluon leading to a Yukawa interaction term in addition to the
confining potential. Clearly, it is a challenge to go beyond the London 
limit  in the sense of taking into account the finite thickness of
vortices and the vanishing of the monopole field inside the vortex core.
One might then naturally ask, whether one gets besides of the
confining potential a Coulomb potential (as used in quarkonium
spectroscopy) instead of the Yukawa one, or possibly something 
between them \cite{baker-1998}. 

\par
The main goal of this paper is an attempt to extend the usual path-integral
approach as much as possible beyond the London limit, paying special
attention to the treatment of boundary terms related to the
nonconfining (shorter range) part of the potential. For this aim and also by
pedagogical reasons, we find it convenient to use throughout
differential form techniques allowing for a transparent and compact
treatment (for definitions, see Table~\ref{tbl:notations1} 
and~\ref{tbl:notations2} ).
As in earlier works (see e.g. 
\cite{tsuzuki-1988,suganuma-1995-npb,
 akhmedov-1995,antonov-1998}), the interaction with
an external $q\bar q$-pair is introduced into the dual field strength
$F$ of
the DAH model in the form of an (open) external electric Dirac string 
$\Sigma^{\mathrm{open}}$ as,
$F = dB - 2\pi * \Sigma^{\mathrm{open}},$
so that the dual Bianchi identity is broken, $dF\neq 0$.
Differing from the above quoted papers, we find it, however, convenient   
to decompose the dual gauge potential $B$ into a regular and a singular
part 
\cite{ball-1988,baker-1991,koma-2000}
so that the singular part cancels the Dirac
string in $dF$ leaving a Coulomb term which keeps the broken dual
Bianchi identity intact. Moreover, for performing 
the path-integration
over the KR field, special emphasize has to be paid to the corresponding
gauge condition. 
As our analysis shows, for a rough estimate based on a 
space time ``mesh size''
larger than the coherence length $m_{\chi}^{-1}$ (chosen as the inverse
of the effective cut-off in the momentum integrals),
the London limit
estimate of the effective string action resulting from field 
distributions outside the string core remains
approximately valid. Clearly, in this case, we cannot see
the inside of the ANO vortex, whose contribution has to be estimated
separately by using classical field equations of motions. Below we only
quote this expression without doing a numerical estimate (this would
first require to fix the DAH parameters from lattice data).
The main result of the paper is an effective string action from which
one gets a string tension given as a sum of a core
and a gauge field (``vortex surface'') contribution, and a nonconfining 
potential having the form of a modified Yukawa interaction.
Obviously, a more complete study of the DAH model,
as considered here, would require the inclusion of
quantum fluctuations of the
string~\cite{gervais-1975,luscher-1980,baker-2000}
 which is, however, outside the scope of this 
analysis.

\par
The organization of the paper is the following.
In Sec.~\ref{sec:DAH} we formulate the DAH model in differential forms
using a field decomposition with cancellation of the Dirac string
in the dual field strength.
In Sec.~\ref{sec:Path} the path-integral approach is discussed paying special
attention to the gauge fixing condition for the KR field and boundary
terms related to nonconfining electric current interactions.
Sec.~\ref{sec:String} is devoted to the derivation 
of the effective action, and Sec.~\ref{sec:Summary}
contains conclusions.

\section{The dual Abelian Higgs model}\label{sec:DAH}

\par
In this section, we formulate the DAH model using differential
form techniques  (see, Tables~\ref{tbl:notations1} and~\ref{tbl:notations2}
)~\cite{nakahara-diffform}.
Let the dual gauge field and the complex scalar
monopole field be $B$ ($1$-form) and $\chi= \phi \exp (i\eta)$
($0$-form),
respectively, then,
the DAH model with an external electric Dirac string
$\Sigma^{\mathrm{open}}$,
whose ends are electric charges \( q \) and \( \bar{q} \),
in Euclidean space-time is given by 
\be
S(B,\chi,\Sigma^{\mathrm{open}}) = \frac{\beta_{g}}{2}(F)^{2}
+(d\phi)^{2}+((B+d\eta)\phi)^{2}
+\lambda(\phi^{2}-v^{2})^{2},
\label{eq:DAHoriginal}
\ee
where the dual field strength  $F$ is expressed as
\be
F = dB - 2\pi * \Sigma^{\mathrm{open}}.
\label{eq:dSigma}
\ee
Due to the presence of the electric Dirac string $\Sigma^{\rm open}$
($2$-form), the dual Bianchi identity is broken as
\be
dF=-2\pi d *\Sigma^{\rm open}=-2\pi *  \delta\Sigma^{\rm open}=
2 \pi * j \ne 0,
\label{eq:dF}
\ee
where  the relation $\delta \Sigma^{\mathrm{open}} = - j$
is used.
This relation just shows that  $\Sigma^{\rm open}$ is nothing else but the
world sheet of the electric Dirac string
whose boundary is the electric current $j$ (1-form).
Clearly, if there is no external electric current,
one must set $\Sigma^{\rm open}=0$.
The inverse of the dual gauge coupling is denoted by $\beta_{g}=1/g^{2}$,
the strength of the self-interaction of the monopole field by $\lambda$,
and the monopole condensate by $v$.
These couplings are related to the mass of the dual gauge boson
and the monopole mass as
$m_{B} \equiv \sqrt{2/\beta_{g}} v = \sqrt{2} g v$ and
$m_{\chi}=2\sqrt{\lambda}v$, which determine  not only the type
of the superconductor vacuum through the so-called Ginzburg-Landau
(GL) parameter
$\kappa= m_{\chi}/m_{B}$, but also the thickness of the flux tube
when the classical solution is considered.
The value $\kappa < 1$ ($\kappa  >1$) describes the type-I (type-II)
vacuum.
Note that the DAH model is invariant under the transformation of
fields
$\chi \mapsto \chi \exp (i\theta)$, $B \mapsto B - d \theta$,
when the U(1) dual gauge symmetry is not spontaneously broken.

\begin{table}[tp]
\centering
\caption{Definitions in differential forms in the four-dimensional 
Euclidean space-time.}
\begin{tabular}{lll}
\hline
$r$-form ($0\le r \le 4$)& $\omega$ &$\omega \equiv   
\frac{1}{r!} \omega_{\mu_1\ldots\mu_r} 
dx_{\mu_1} \wedge \cdots\wedge dx_{\mu_r}$
\\ 
\hline
exterior derivative & $d$ &
 $r$-form $\mapsto$ $(r+1)$-form 
\\
Hodge star & $*$ &
$r$-form $\mapsto$ $(4-r)$-form
\\
&$**$&multiply a factor $(-1)^{r}$ for $r$-form
\\
codifferential &$\delta \equiv - *d*$&
$r$-form $\mapsto$ $(r-1)$-form 
\\
Laplacian &$\Delta \equiv d\delta + \delta d$&
$r$-form $\mapsto$ $r$-form
\\
Inner product & $(\omega,\eta)\equiv \int \omega \wedge * \eta$ &
$(\omega,\eta)=
\frac{1}{r!}\int d^{4}x \omega_{\mu_1\ldots\mu_r} 
\eta_{\mu_1\ldots\mu_r} \quad$  ($\omega, \eta \in r$-form)
\\
&&$(\omega)^{2} \equiv (\omega,\omega)$
\\
\hline
\end{tabular}
\label{tbl:notations1}
\end{table}

\begin{table}[tp]
\centering
\caption{Ingredients of the DAH model in the differential-form notation.}
\begin{tabular}{lrl}
\hline
dual gauge field & 
$1$-form &
$B\equiv B_{\mu}dx_{\mu}$ \\
monopole field  &
$0$-form &
$\chi\equiv \phi \exp (i \eta)$ \\
electric Dirac string &
$2$-form &
$\Sigma \equiv \frac{1}{2}\Sigma_{\mu\nu} dx_{\mu} \wedge dx_{\nu}$
\\
electric current &
$1$-form &
$j \equiv j_{\mu} dx_{\mu}$
\\
\hline
\end{tabular}
\label{tbl:notations2}
\end{table}

\par
The electric Dirac world sheet singularity which explicitly appears
in the dual field strength~\eqref{eq:dSigma} has the standard
form and is required
to satisfy the broken Bianchi identity~\eqref{eq:dF}. 
Clearly, such a
singularity would give a divergent contribution in $(F)^2$ and must
therefore be cancelled by a corresponding singular term in $dB$. 
Thus, it is useful to decompose the dual gauge field into two parts,
the regular quantum part not containing an electric Dirac string and
the singular part with an electric Dirac string, as
\be
B=B^{\rm reg}+B^{\rm sing},
\label{eq:decompB}
\ee
where the singular part has the explicit form
\be
B^{\rm sing} \equiv 2\pi \Delta^{-1}\delta * \Sigma^{\rm open}.
\ee
Here the inverse of the Laplacian,
$\Delta^{-1}$, is the Coulomb propagator.
Then, by using the relation $d \Delta^{-1} \delta
+\delta\Delta^{-1}d=1$ and the equation 
$\delta \Sigma^{\rm open}= -j$, we have
\be
dB^{\rm sing}=
2\pi *\Sigma^{\rm open}+2\pi \Delta^{-1}\delta*j
= 2\pi (*\Sigma^{\rm open} +  * C),
\label{eq:dBs}
\ee
where $*C$ is the $2$-form field
\be
*C =  \Delta^{-1}\delta*j .
\ee
The dual field strength is, then, written as%
\footnote{Note that the equality of $F$ in 
Eqs.~\eqref{eq:dSigma}, \eqref{eq:ourF} does not at
all mean that the Dirac string world sheet is simply replaced by the
Coulomb electric field of the quark charges, since both objects are
combined with different fields $B$ and $B^{\rm reg}$ having different
boundary conditions.  After the decomposition~\eqref{eq:decompB} 
the Dirac string explicitly appears in the interaction term 
of Eq.~\eqref{eq:DAHours},
where it dictates the boundary condition of the monopole 
field $\phi$ which has to vanish at the string core.}
\be
F = dB^{\rm reg} + 2\pi *C.
\label{eq:ourF}
\ee
In the $q\bar{q}$ system, $*C$ turns out to contain the Coulomb
electric field originating from the electric charges.
In fact, we have
\be
d*C = \Delta^{-1}d \delta*j = \Delta^{-1} \Delta * j
= *j ,
\label{eq:dCdelC}
\ee
where we have used the electric current
conservation condition, \( \delta j = 0 \).
Note that the dual Bianchi identity for the dual gauge field
$d^{2}B=0$ is,
of course, satisfied even after the decomposition into the
regular and the singular parts, since we have
$d^{2}B^{\rm reg}=0$ and $d^{2} B^{\rm sing}=2\pi
(d * \Sigma^{\rm open} + d * C)=2\pi(-*j + *j)=0$.
Thus we still have the relation $dF=2\pi d * C = 2\pi *j$
as in Eq.~\eqref{eq:dF}.
Using the relation $\delta^{2}=0$ one can further show $\delta * C = 0$.

\par
In the case that the phase of the monopole field is singular (multivalued),
we also have closed electric Dirac strings $\Sigma^{\rm closed}$.
This structure becomes manifest, if we write the phase
with the regular and singular parts as
\be
\eta = \eta^{\rm reg} +\eta^{\rm sing}, \\
\ee
where each part is defined so as to satisfy the relation
\be
d^{2} \eta^{\rm reg}=0, \quad d^{2} \eta^{\rm sing}
= 2\pi *\Sigma^{\rm closed} \ne 0.
\ee
Such a closed world sheet singularity can be regarded as the origin of
a glueball excitation \cite{koma-1999}.
However, since this is not the issue of  interest here,
we neglect such  singular phase contributions assuming that the phase
is single-valued.
Thus, from hereafter we simply use $\Sigma$ as the world sheet of
the open electric Dirac string, omitting the explicit label ``open''.
The DAH action can then be written  as
\be
S (B,\chi,\Sigma) = \frac{\beta_{g}}{2}(dB^{\rm reg}+2 \pi * C )^{2}
+(d\phi)^{2}
+((B^{\rm reg}+B^{\rm sing}+d\eta^{\rm reg})\phi)^{2}
+\lambda(\phi^{2}-v^{2})^{2}.
\label{eq:DAHours}
\ee

\section{Path integral transformation to Kalb-Ramond 
fields}\label{sec:Path}

\par
In this section, in order to obtain the string representation,
we shall next perform a field transformation in the path integral 
representation
of the partition function of the DAH model given by
\bea
{\cal Z}(\Sigma) &=&
\int {\cal D}B^{\rm reg}
\delta [ \delta B^{\rm reg}-f_{B}]
\phi {\cal D}\phi {\cal D}\eta^{\rm reg}
\exp
\left [ -S(B,\chi,\Sigma) \right ],
\label{eq:Zsigmastart}
\eea
where the DAH action has the form quoted in Eq.~\eqref{eq:DAHours}.
In the integral measure, we have inserted a usual
gauge fixing term for the regular part of the dual gauge field
$B^{\rm reg}$.
The corresponding Faddeev-Popov (FP) determinant is omitted,
since it contributes a trivial constant factor in an Abelian theory.
We start from the linearization of the square term
$ ( (B^{\rm reg}+B^{\rm sing}+d\eta^{\rm reg}) \phi  )^{2}$
by means of a  $1$-form auxiliary field $E$ as
\bea
&&
\exp \left [ -\left (
(B^{\rm reg}+B^{\rm sing}+d\eta^{\rm reg}) \phi
\right )^{2}\right ]
\nonumber\\*
&&
=
\phi^{-4} \int {\cal D}E \exp
\left [ -\left \{
  (E,\frac{1}{4\phi^{2}}E)
- i (E, B^{\rm reg}+B^{\rm sing}+d\eta^{\rm reg})
\right \}
\right ],
\eea
so that the integration measure of the modulus of the monopole field
$\phi$ in Eq.~\eqref{eq:Zsigmastart} is modified as $\phi {\cal D}\phi
\times \phi^{-4} \to \phi^{-3} {\cal D}\phi$.
Then, based on the relation $(E,d\eta^{\rm reg})
= (\delta E,\eta^{\rm reg})$, we can integrate over the
regular part of the phase $\eta^{\rm reg}$, which leads to
the delta functional $\delta [ \delta E ]$ in the integration measure.
The constraint on $E$ can be resolved by introducing  the
2-form Kalb-Ramond (KR) field $h$ as
\be
\delta [ \delta E ] = \int {\cal D}h \delta [ \delta h -f_{h}]
\delta [ E- \delta * h ],
\ee
where the so-called hyper-gauge fixing delta functional appears
in order to avoid the overcounting in the integration over $h$, which
is due to the hyper-gauge invariance $h \mapsto h + d \Lambda$
with 1-form field $\Lambda$. The corresponding FP determinant
is again omitted due to the same reason as for the dual gauge field.
Now, we can immediately perform the integration over the
auxiliary field $E$ as
\bea
&&{\cal Z}(\Sigma) =
\int {\cal D}B^{\rm reg}
\delta [ \delta B^{\rm reg}-f_{B}]
\phi^{-3}{\cal D}\phi
{\cal D}h \delta [\delta h - f_{h}]
\nonumber\\*
&&\times \exp
\Biggl [ - \Biggl \{
\frac{\beta_{g}}{2}(dB^{\rm reg}+2\pi * C)^{2}
+(d\phi)^{2}
+ (\delta *h,\frac{1}{4 \phi^{2}}\delta *h )
\nonumber\\*
&&
-i(\delta*h, B^{\rm reg}+B^{\rm sing})
+\lambda(\phi^{2}-v^{2})^{2}
\Biggr \} \Biggr ].
\eea
Here, the kinetic term of the dual gauge field 
can be further rewritten as%
\footnote{Here and in other cases of partial integration, 
arising surface integrals 
are vanishing due to the vanishing of the regular 
fields at infinity.}
\be
(dB^{\rm reg}+2\pi * C)^{2} = (dB^{\rm reg})^{2} + 4\pi^{2}(j,
\Delta^{-1} j) ,
\ee
where the cross term $(dB^{\rm reg},* C)=(B^{\rm reg}, \delta *C)$ 
vanishes due to the fact that $\delta * C=0$.
Moreover, other terms in the DAH action are also rewritten as
\bea
(* C)^{2}
&=& (\Delta^{-1}\delta * j,\Delta^{-1}\delta * j)
= (j, \Delta^{-1} j ),\\
(\delta *h,\frac{1}{4 \phi^{2}}\delta *h )
&=& (dh,\frac{1}{4\phi^{2}}d h),\\
(\delta*h, B^{\rm reg}+B^{\rm sing}) &=&
(\delta*h, B^{\rm reg}) +2  \pi (h,\Sigma)
+  2\pi (\delta h, \Delta^{-1}j),
\eea
where the relation, $(*h, * C) = (\delta h,\Delta^{-1}j)$, has
been taken into account.
The partition function is then written as
\bea
{\cal Z}(\Sigma) &=&
\int {\cal D}B^{\rm reg} \delta [ \delta B^{\rm reg}-f_{B}]
\phi^{-3}{\cal D}\phi
{\cal D}h \delta [\delta h - f_{h}]
\nonumber\\*
&&
\times \exp \Biggl [ -\Biggl \{
\frac{\beta_{g}}{2} (dB^{\rm reg})^{2}+ 2 \pi^{2} \beta_{g}
(j,\Delta^{-1} j)
+(d\phi)^{2} + (dh, \frac{1}{4\phi^{2}}dh)
-i(\delta*h, B^{\rm reg})
\nonumber\\*
&& \quad
-2 \pi i (h, \Sigma)
- 2 \pi i (\delta h,\Delta^{-1}j) +\lambda(\phi^{2}-v^{2})^{2}
\Biggr \}
\Biggr ].
\label{eq:DAHrewrite}
\eea
Next, the integration over the regular part of the dual gauge field $B^{\rm
reg}$
is achieved in a standard way by an insertion of the identity
\be
const.
= \int {\cal D}f_{B}
\exp \left [ -\frac{\beta_{g}}{2\xi_{B}} f_{B}^{2}\right ].
\label{eq:Bfix}
\ee
Performing the integration over $f_{B}$ and taking the
Landau gauge $\xi_{B}=1$, we get the terms
\bea
\frac{\beta_{g}}{2} (B^{\rm reg},\Delta B^{\rm reg})
-i(\delta*h, B^{\rm reg})
&=&
\frac{\beta_{g}}{2}
\left ( 
\{B^{\rm reg}-\frac{i}{\beta_{g}}\Delta^{-1}\delta*h \},
\Delta
 \{B^{\rm reg}-\frac{i}{\beta_{g}}\Delta^{-1}\delta*h \}
 \right )
\nonumber\\*
&&
+\frac{1}{2\beta_{g}} \left (\delta *h, \Delta^{-1} \delta *h \right
),
\eea
where the last term can be rewritten as
\bea
(\delta *h ,\Delta^{-1}\delta *h  )
&=&
(h)^{2} -(\delta h,  \Delta^{-1}\delta h  ).
\eea
Then, the Gaussian integration over the shifted dual gauge field,
$B^{\rm reg}-\frac{i}{\beta_{g}}\Delta^{-1}\delta*h  \to B^{\rm reg}$,
leads to the expression
\bea
{\cal Z}(\Sigma) &=& \int
\phi^{-3}{\cal D}\phi {\cal D}h \delta [\delta h - f_{h}]
\exp
\Biggl [ - \Biggl \{ 2 \pi^{2} \beta_{g} (j,\Delta^{-1} j)
+(d\phi)^{2} + (dh, \frac{1}{4\phi^{2}}dh)
+  \frac{1}{2\beta_{g}} (h)^{2} 
\nonumber\\*
&&
- \frac{1}{2\beta_{g}}(\delta h,
\Delta^{-1}\delta h  )
-2 \pi i (h, \Sigma)
- 2 \pi i (\delta h,\Delta^{-1}j)
+\lambda(\phi^{2}-v^{2})^{2}
\Biggr \}
\Biggr ].
\eea

\section{String representation}\label{sec:String}

\par
In this section, we aim to clarify the structure of the
string representation of the DAH model.
First, we divide the action into three parts as
\be
S = S^{(1)} + S^{(2)} + S^{(3)},
\ee
where each action is defined by
\bea
S^{(1)} &=& 2 \pi^{2} \beta_{g} (j,\Delta^{-1} j),
\label{eq:action-Coulomb}\\
S^{(2)} &=&
(d\phi)^{2}
+  (dh, \frac{1}{4}\{\frac{1}{\phi^{2}} -\frac{1}{v^{2}}\}  dh)
+ \lambda(\phi^{2}-v^{2})^{2},
\label{eq:action-core}\\
S^{(3)} &=&
\frac{1}{4v^{2}}(dh)^{2}
+\frac{1}{2\beta_{g}} (h)^{2}
- \frac{1}{2\beta_{g}}(\delta h, \Delta^{-1}\delta h  )
-2 \pi i (h, \Sigma) - 2 \pi i (\delta h,\Delta^{-1}j),
\label{eq:action-surface}
\eea
respectively.
The first term $S^{(1)}$ leads to the pure
Coulomb potential (pure boundary contribution)
when the static quark-antiquark  system 
is  investigated.
The second term $S^{(2)}$ is defined so as to give a zero contribution
to the effective string action in the case that $\phi =v$,
which  usually corresponds to taking the London limit
$\lambda \to  \infty$.
In other words, this term leads for finite $\lambda$ to a nonvanishing
contribution due to the finite thickness (size of the core) of the
string modelling the flux tube (ANO vortex), inside which the modulus of
the monopole field smoothly becomes zero, 
$\phi=0$.
The third term $S^{(3)}$ is mainly responsible for 
the field contributions outside the core, near
the  surface of the flux tube, which remains even in the case
that the monopole modulus has a constant value, $\phi=v$.
It is interesting to evaluate the DAH action 
just on the string world sheet  by taking into account 
the boundary conditions of the classical field equation.
One finds that in order to get a finite energy contribution, we need to 
impose both $\phi=0$ and $dh=0$ on the string world sheet,
where the second condition is 
resolved by $h=d A$ with 1-form field $A$.
Then, by inserting this into Eq.~\eqref{eq:action-surface}, 
we see that the contribution to the $S^{(3)}$ from the string world sheet 
is zero.

\par
In order to discuss the effective string action definitely, let us
consider the case that $m_{B} < m_{\chi}$.  For the rough space-time
structure, whose mesh size is larger than $m_{\chi}^{-1}$, the London
limit picture based on a dominating expression $S^{(3)}_{>
m_{\chi}^{-1}}$ is valid, where the subscript means that one has to
integrate over transverse distances from the string, $\rho >
m_{\chi}^{-1}$, with the monopole mass $m_{\chi}$ chosen as an
effective cut-off $\Lambda_{\rm eff}$. 
 We will discuss the arbitrariness of choice of this 
effective cut-off later.
Clearly, in this case, we cannot see the inside of
the flux tube.  On the other hand, to see the finer structure of the
flux tube, variations of the monopole field in $S^{(2)}$ should be
taken into account.  
The effective action of the vortex ``core''
contribution is described by $S^{(2)} + S^{(3)}_{< m_{\chi}^{-1}}$,
and its leading term contains
the Nambu-Goto action with the string tension
$\sigma_{\rm core}$ and a current term $S_{\mathrm{core}}(j)$, 
\be
S_{\rm core} = S_{\mathrm{core}}(j)+\sigma_{\rm core} \int
d^{2}\xi \sqrt{g(\xi)} , 
\ee
where $\xi^{a}$ $(a=1,2)$ parametrize the string world
sheet described by the coordinate $\tilde{x_{\mu}}(\xi)$, and $g(\xi)$
is the determinant of the induced metric, $g_{ab}(\xi) \equiv
\frac{\tilde{x}_{\mu}(\xi)}{\partial \xi^{a}}
\frac{\tilde{x}_{\mu}(\xi)}{\partial \xi^{b}}$. 
The string tension
$\sigma_{\rm core}$ is controlled by the solution of field equations
derived from the action in the core region~\cite{forster-1974,
davis-1988}.  Note that $S_{\mathrm{core}}(j)$ results only from
$S^{(3)}_{< m_{\chi}^{-1}}$ \footnote{ If one approximately considers
a vortex core with radius $m_{\chi}^{-1}$ in which $\phi=0$ and $dh=0$
everywhere as on the string world sheet, one finds
$S_{\mathrm{core}}(j)=0$, since $S^{(3)}_{< m_{\chi}^{-1}}=0$.}.

\par
Let us evaluate the  string effective action of the ``surface
contribution''
described by $S^{(3)}_{> m_{\chi}^{-1}}$.
To do this, we first integrate out the KR field, and then
extract the surface contribution from it by taking into
account a suitable regularization in transverse variables.
Since the action $S^{(3)}$ does not depend on the monopole modulus \( \phi \),
the corresponding partition function is written as
\bea
&&
{\cal Z}^{(3)}
=
\int {\cal D}h \delta [\delta h-f_{h} ]
\nonumber\\*
&&
\times
\exp \Biggl [
- \Biggl \{
4 \pi^{2}v^{2}  \left( \Sigma, D\Sigma\right) 
+\frac{1}{4 v^{2}}
\left ( \{h - 4 \pi v^{2} i D \Sigma\},
D^{-1}
\{h - 4 \pi v^{2} i D \Sigma \}
\right )
\nonumber\\*
&&
-2\pi^{2}\beta_{g} \left ( j,
\left \{ D - \Delta^{-1} \right \}
j \right )
- \frac{1}{4v^{2}}
\left (\{\delta h +4 \pi v^{2} i Dj \},
 D^{-1} \Delta^{-1}
\{\delta h+ 4 \pi v^{2} i Dj \}\right )
\Biggr \} \Biggr ].
\label{eqn:integ-KR}
\eea
where we have defined the propagator of the massive KR field
 $D \equiv (\Delta + m_{B}^{2})^{-1}$ and used the relation
$(dh)^{2} = (h, \Delta h)-(\delta h )^{2}$.
The integration over the KR field is achieved in a similar way as for
the dual gauge field by inserting an identity in form of a
path-integral over the hyper-gauge
fixing function $f_{h}$,
\be
const. =
\int {\cal D}f_{h}
\exp
\Biggl [
-  \frac{1}{4 v^{2}\xi_{h}}
\left (\{f_{h}+4 \pi v^{2} i Dj \},
D^{-1} \Delta^{-1}
\{f_{h}+4 \pi v^{2} i Dj\}
\right )
\Biggr ].
\ee
Note that the integration over $f_{h}$ and taking the hyper-Landau gauge
$\xi_{h}=1$, leads to a cancellation of
the last term of the action in the partition function
\eqref{eqn:integ-KR}.
Then, we can integrate over the shifted KR field through the
replacement
$h - 4 \pi v^{2} i D \Sigma \to h$.
The resulting effective string action from the surface contribution is
then given by
\bea
S^{(3)}_{>m_{\chi}^{-1}}
&=&
   4 \pi^{2}v^{2}
\left( \Sigma, D \Sigma \right)
\biggr |_{>m_{\chi}^{-1}}
+2\pi^{2}\beta_{g} \left ( j,\left \{    D  - \Delta^{-1} \right \} j  \right )
\biggr |_{>m_{\chi}^{-1}},
\label{eq:outfinal}
\eea
where ``$|_{>m_{\chi}^{-1}}$'' means that
a corresponding effective cutoff (mesh size) should be taken into account.
One finds that the first term represents the interaction between
world sheet elements of the electric Dirac string via the
propagator of the massive KR field.
In tensor form, this expression can be written as
\be
\left( \Sigma, D \Sigma \right) \biggr|_{>m_{\chi}^{-1}}
=
\frac{1}{2}\int_{}^{} d^{4}x \int_{}^{} d^{4}y
\Sigma_{\mu\nu}(x) D(x-y)
\Sigma_{\mu\nu}(y)
\biggr |_{>m_{\chi}^{-1}}
\label{eq:outactiontensor}
\ee
where
\be
\Sigma_{\mu \nu}(x)
=
\int_{\Sigma} d^{2}\xi \sqrt{g(\xi)} t_{\mu \nu}(\xi)
\delta^{(4)}(x - \tilde{x}(\xi)) ,
\label{eq:defsigma}
\ee
and $t_{\mu\nu}(\xi) = \frac{\epsilon^{ab}}{\sqrt{g(\xi)} }
\frac{\partial \tilde{x}_{\mu}(\xi)}{\partial \xi^{a}}
\frac{\partial \tilde{x}_{\nu}(\xi)}{\partial \xi^{b}}$ is an
antisymmetric tensor which determines the orientation of the string
world  sheet $\Sigma$.
It is important to note that the regularization is achieved by
introducing transverse coordinates $\Xi^{k}$ $(k=3,4)$, which
parametrize the  direction  d
perpendicular  to the string world sheet~\cite{orland-1994}. 
In other words, points in Euclidean space close enough to $\Sigma$ are
parametrized as $x=x(\xi,\Xi)$ ~\cite{orland-1994,gervais-1975}.

Then the massive KR propagator in Eq.~(\ref{eq:outactiontensor})
can be written as
\begin{eqnarray}
       D(x-y)& = & (\Delta + m^{2})^{-1}\delta^{(4)}(x(\xi,\Xi)-y(\xi',\Xi'))
\nonumber      \\*
      & = &\left( -\sum_{k=3,4} \frac{\partial^{2}}{\partial \left(
           \Xi^{k} \right)^{2}} + \Delta_{\xi} + m^{2} \right)^{-1}
      \frac{1}{\sqrt{g(\xi)}}\delta^{(2)}(\xi -\xi')
      \delta^{(2)}(\Xi -\Xi'),
     \label{eq:newD}
\end{eqnarray}
where the Laplacian on the string world sheet is defined by
\be
\Delta_{\xi} = -\frac{1}{\sqrt{g(\xi)}} \partial_{a} g^{ab}(\xi)
\sqrt{g(\xi)}
\partial_{b}.
\ee
In this scheme, the coherence length of the monopole field
$m_{\chi}^{-1}$ plays the role of an effective cutoff of the $\Xi$ integral.
By the Taylor expansion of the propagator of the KR field, we obtain
the explicit form of the effective string action as
\bea
&&
4 \pi^{2}v^{2}
\left( \Sigma, D \Sigma \right)
\biggr |_{>m_{\chi}^{-1}}
\nonumber\\*
&&
=
\sigma_{\rm surf}
\int_{}^{} d^{2}\xi \sqrt{g(\xi)}
+
\alpha_{\rm surf}
\int d^{2}\xi \;   \sqrt{g(\xi)}  g^{ab}(\xi)
\left( \partial_{a} t_{\mu \nu}(\xi) \right)
  \left( \partial_{b}  t_{\mu \nu}(\xi)\right)
  + {\cal O}( \Delta_{\xi}^{2}).
\label{eq:expansion}
  \eea
Here, the first term represents the Nambu-Goto action
with the string tension
\be
\sigma_{\rm surf}
= \pi v^{2}\ln \frac{m_B^{2} +m_{\chi}^{2}}{m_B^{2}}
= \pi v^{2}\ln \left(1 +\kappa^{2}\right),
\label{eq:ssurf}
\ee
and  the second term  is the so-called rigidity term with
the negative coefficient
\be
\alpha_{\rm surf}
=
\frac{\pi v^{2} }{2}
\left(\frac{1}{m_{\chi}^{2}+m_B^{2}} - \frac{1}{m_B^{2}}\right)
=
- \frac{\pi \beta_{g}}{4} \frac{\kappa^{2}}{1+\kappa^{2}} \qquad (<
0),
\label{eq:asurf}
\ee
where $\kappa=m_{\chi}/m_B$ is the GL parameter.
Note that the rigidity term  appears as a first order contribution
in the derivative expansion with the Laplacian $\Delta_{\xi}$.
The second term of \eqref{eq:outfinal}, which is induced from the
boundary of the string world sheet, is evaluated in a similar 
``effective regularization''
scheme,  discussed below.
Finally, by combining it with the pure Coulomb term $S^{(1)}$,
we get the effective action 
\bea
S_{\mathrm{eff}} (\Sigma)
&=&
S (j)
+(\sigma_{\rm core} + \sigma_{\rm surf}) \int d^{2}\xi \sqrt{g(\xi)}
\nonumber\\*
&&
+ \alpha_{\rm surf}\int d^{2}\xi \;   \sqrt{g(\xi)}  g^{ab}(\xi)
\left( \partial_{a} t_{\mu \nu}(\xi) \right)
  \left( \partial_{b} t_{\mu \nu}(\xi)\right)
+ {\cal O}( \Delta_{\xi}^{2}),
\label{eq:effaction}
\eea
where the boundary (electric current) contribution is given by
\be
S (j) =S_{\mathrm{core}}(j) 
+ \frac{1}{2 \beta_{e}} (j, \Delta^{-1} j) 
+ \frac{1}{2 \beta_{e}}
\left ( j, \left \{    D  - \Delta^{-1} \right \} j  \right )
\biggr |_{>m_{\chi}^{-1}}
\label{effaction-boundary}
\ee
Note that the Dirac quantization condition
$4\pi^{2}\beta_{e} \beta_{g}=1$ $(eg=4 \pi)$ is taken into account,
where $\beta_{e}=4/e^{2}$ and $\beta_{g}=1/g^{2}$.
Eqs.~\eqref{eq:effaction} and \eqref{effaction-boundary} are the
main result of this paper.

\par
Here we would like to mention  the role of the effective cut-off 
$\Lambda_{\rm eff} = m_{\chi}$ for the 
evaluation of  $S^{(3)}$.
In fact, we can choose
any scale to divide into the low and the high energy parts as
$S^{(3)}_{<\Lambda_{\rm eff}^{-1}}$ and 
$S^{(3)}_{>\Lambda_{\rm eff}^{-1}}$.  
If $\Lambda_{\rm eff} = c m_{\chi}$ (where $c \ne 1$) is taken, 
not only the $\sigma_{\mathrm{surf}}$ in Eq.~\eqref{eq:ssurf} 
but also $S_{\mathrm{core}} \equiv S^{(2)} +
S^{(3)}_{<\Lambda_{\rm eff}^{-1}}$ are changed. 
However, the final expressions
\eqref{eq:effaction} and \eqref{effaction-boundary} are not affected by
the choice of $\Lambda_{\rm eff}$,
since the changes in $\sigma_{\mathrm{surf}}$
is absorbed by $\sigma_{\mathrm{core}}$, and the change in the third
term of Eq.~\eqref{effaction-boundary} is absorbed by
$S_{\mathrm{core}}(j)$.
Due to the fact that $S^{(2)}$ contributes only in the region
at $\rho < m^{-1}_{\chi}$, the choice
$\Lambda_{\rm eff} =
m_{\chi}\,\, (c=1)$ for $S^{(3)}$ turns out to be the most 
``effective'' one,  which we take in this paper.

\par
It is interesting to discuss the boundary contributions of the
string world sheet.
In order to get an explicit form, let us evaluate them with the
static electric current
\be
j_{\mu}(x) = \delta_{\mu 0} \{ \delta^{(3)}(\bvec{x}-\bvec{a})
-\delta^{(3)}(\bvec{x}-\bvec{b}) \},
\ee
where $\bvec{a}$ and $\bvec{b}$ are the positions  of the 
electric charges (The charge $e$ is already factorized out).
Denoting the distance between electric charges as
$r=|\bvec{a}-\bvec{b}|$, the static potential is given by
\bea
V(r) &=& V_{\mathrm{core}}(r)
-\frac{1}{4\pi \beta_{e}r}
\nonumber\\*
&&
+
\frac{1}{2 \beta_{e}}
\int_{p_{\rho} < m_{\chi}}
\frac{d^{3}p}{ (2\pi)^{3}}
\left ( 1 -e^{-i \bvec{p} \cdot \bvec{r}} \right )
\left ( 1 -e^{  i \bvec{p} \cdot \bvec{r}} \right )
\left [
\frac{1}{\bvec{p}^{2}+m_{B}^{2}}
- \frac{1}{\bvec{p}^{2}}
\right ],
\eea
where
$p_{\rho}$  is the momentum in the transverse direction,
perpendicular to $\bvec{r}$.
In the rough approximation 
that $\phi = 0$ in the whole core region (it also means $dh=0$ 
in the core),
we have $V_{\mathrm{core}} = 0 $, 
due to the fact that $S_{\rm core} (j)=0$,
and we get the final expression for the potential
\be
V(r) =
-\frac{e^{-m_{B}r}}{4\pi \beta_{e}r}
\left [
1- e ^{-(\sqrt{ m_{\chi}^{2}+m_{B}^{2} }-m_{B})r}
+ e ^{-(m_{\chi} -m_{B}) r}
\right ],
\ee
where constant terms have been dropped.
Clearly, this form, which is valid for 
$r>1/m_{\chi}$, is not the pure Yukawa potential nor a Coulomb
potential.
However, it is interesting to note that in the London limit $m_{\chi}
\to \infty$ $(\lambda \to \infty)$, the potential reproduces
the usual Yukawa potential.
The complete potential includes, of course, the confining potential 
$V_{\mathrm{conf}}=(\sigma_{\mathrm{core}}+\sigma_{\mathrm{surf}}) r$
arising from the Nambu-Goto 
action in \eqref{eq:effaction}.

\section{Summary and conclusions}\label{sec:Summary}

\par
The present paper is a first attempt to extend earlier path-integral
investigations of the string representation of the DAH model
\cite{orland-1994,sato-1995,akhmedov-1995, 
antonov-1998}
 as much as possible beyond the London limit.
Particular attention was given to the treatment of boundary terms
related to the nonconfining part of the $q\bar q$-potential.
In fact, for a vortex with finite thickness and a vanishing monopole field
inside the core, the usual expression for the Yukawa potential is
expected to become modified. In order not to exclude from the very
beginning even the possible appearance of Coulomb interactions (as
indicated in phenomenological applications), we found it convenient to use a 
particular field decomposition, where the standard Dirac string
describing the external quark-antiquark source is cancelled just
keeping a Coulomb term which satisfies the broken dual Bianchi identity. 
For the proper treatment of boundary terms, we found it very
convenient to use differential form techniques which, after performing
a transformation to antisymmetric KR fields, required a careful
treatment of corresponding gauge conditions for performing necessary
path integrations. The investigation of the effective action requires the
introduction of a  cutoff. For this aim, analogously to
Refs.~\cite{orland-1994,gervais-1975}, we reparametrized the integration
variables $x,y$ in Eq.~\eqref{eq:outactiontensor}
near the string world sheet $\Sigma$
in coordinates longitudinal
and orthogonal to it, and used the monopole
mass $m_{\chi}$ as an effective 
cutoff in the resulting transverse momentum integrals.
The derivative expansion of the effective action then leads to the
Nambu-Goto action and a rigidity term, with expressions for the (finite)
string tension and the negative rigidity coefficient formally close
to those of the London  limit. Obviously, the string tension now gets an
additional contribution from the vortex core which has to be
calculated by using the classical field equations. 
Finally, concerning the nonconfining potential, there arises in the chosen
regularization an interesting cancellation of the Coulomb term, 
originally appearing in the dual field strength in 
Eq.~\eqref{eq:ourF}, 
by a corresponding term of the boundary
contribution of the KR field leaving instead a  modified Yukawa interaction.
In conclusion, we remark that the extension of these investigations to the
more realistic $SU(3_c)$-DAH model is now under further investigation.

\begin{acknowledgments}
The authors thank M.I.~Polikarpov  and 
T.~Suzuki for very fruitful discussions.
One of the authors (D.E.) also thanks  N. Brambilla for interesting 
discussion concerning the role of the London 
limit.
D.E. acknowledges the support provided to him by
the Ministry of Education and Science and Technology of Japan (Monkasho).
One of the authors (Y.K.) is partially supported  by 
the Ministry of Education, Science, Sports and Culture,
Japan, Grant-in-Aid for Encouragement of Young 
Scientists (B), 14740161, 2002.
\end{acknowledgments}


\begin{thebibliography}{10}
\expandafter\ifx\csname bibnamefont\endcsname\relax
  \def\bibnamefont#1{#1}\fi
\expandafter\ifx\csname bibfnamefont\endcsname\relax
  \def\bibfnamefont#1{#1}\fi
\expandafter\ifx\csname url\endcsname\relax
  \def\url#1{\texttt{#1}}\fi
\expandafter\ifx\csname urlprefix\endcsname\relax\def\urlprefix{URL }\fi
\expandafter\ifx\csname bibinfo\endcsname\relax \def\bibinfo#1#2{#2}\fi
\expandafter\ifx\csname eprint\endcsname\relax \def\eprint#1{#1}\fi



\bibitem{tHooft-1976}
\bibinfo{author}{\bibfnamefont{G.}~\bibnamefont{'t~Hooft}},
  in \emph{\bibinfo{title}{{High Energy Physics}}}
  (\bibinfo{publisher}{Editrice Compositori}, \bibinfo{address}{Bologna},
  \bibinfo{year}{1976}).

\bibitem{mandelstam}
\bibinfo{author}{\bibfnamefont{S.}~\bibnamefont{Mandelstam}},
  \bibinfo{journal}{Phys. Rept.} \textbf{\bibinfo{volume}{23}},
  \bibinfo{pages}{245} (\bibinfo{year}{1976}).

\bibitem{abrikosov-1957}
\bibinfo{author}{\bibfnamefont{A.A.}~\bibnamefont{Abrikosov}},
  \bibinfo{journal}{Sov. Phys. JETP} \textbf{\bibinfo{volume}{5}},
  \bibinfo{pages}{1174} (\bibinfo{year}{1957}).

\bibitem{nielsen-1973}
\bibinfo{author}{\bibfnamefont{H.B.}~\bibnamefont{Nielsen}} \bibnamefont{and}
  \bibinfo{author}{\bibfnamefont{P.}~\bibnamefont{Olesen}},
  \bibinfo{journal}{Nucl. Phys.} \textbf{\bibinfo{volume}{B61}},
  \bibinfo{pages}{45} (\bibinfo{year}{1973}).

  \bibitem{nambu-1974}
\bibinfo{author}{\bibfnamefont{Y.}~\bibnamefont{Nambu}},
  \bibinfo{journal}{Phys. Rev.} \textbf{\bibinfo{volume}{D10}},
  \bibinfo{pages}{4262} (\bibinfo{year}{1974}).
  
\bibitem{tsuzuki-1990}
\bibinfo{author}{\bibfnamefont{T.}~\bibnamefont{Suzuki}} \bibnamefont{and}
  \bibinfo{author}{\bibfnamefont{I.}~\bibnamefont{Yotsuyanagi}},
  \bibinfo{journal}{Phys. Rev.} \textbf{\bibinfo{volume}{D42}},
  \bibinfo{pages}{4257} (\bibinfo{year}{1990}).

\bibitem{brandstater-1991}
\bibinfo{author}{\bibfnamefont{F.}~\bibnamefont{Brandst\"ater}},
  \bibinfo{author}{\bibfnamefont{U.J.}~\bibnamefont{Wiese}}, \bibnamefont{and}
  \bibinfo{author}{\bibfnamefont{G.}~\bibnamefont{Schierholz}},
  \bibinfo{journal}{Phys. Lett.} \textbf{\bibinfo{volume}{B272}},
  \bibinfo{pages}{319} (\bibinfo{year}{1991}).

\bibitem{kronfeld-1987}
\bibinfo{author}{\bibfnamefont{A.S.}~\bibnamefont{Kronfeld}},
  \bibinfo{author}{\bibfnamefont{G.}~\bibnamefont{Schierholz}},
  \bibnamefont{and} \bibinfo{author}{\bibfnamefont{U.J.}~\bibnamefont{Wiese}},
  \bibinfo{journal}{Nucl. Phys.} \textbf{\bibinfo{volume}{B293}},
  \bibinfo{pages}{461} (\bibinfo{year}{1987}).

\bibitem{tsuzuki-1988}
\bibinfo{author}{\bibfnamefont{T.}~\bibnamefont{Suzuki}},
  \bibinfo{journal}{Prog. Theor. Phys.} \textbf{\bibinfo{volume}{80}},
  \bibinfo{pages}{929} (\bibinfo{year}{1988});
    \textit{ibid.}
  \textbf{\bibinfo{volume}{81}},
  \bibinfo{pages}{752} (\bibinfo{year}{1989});
\bibinfo{author}{\bibfnamefont{T.}~\bibnamefont{Suzuki}} \bibnamefont{and}
  \bibinfo{author}{\bibfnamefont{S.}~\bibnamefont{Maedan}},
  \textit{ibid.}
  \textbf{\bibinfo{volume}{81}},
  \bibinfo{pages}{229} (\bibinfo{year}{1989}).


\bibitem{suganuma-1995-npb}
\bibinfo{author}{\bibfnamefont{H.}~\bibnamefont{Suganuma}},
  \bibinfo{author}{\bibfnamefont{S.}~\bibnamefont{Sasaki}}, \bibnamefont{and}
  \bibinfo{author}{\bibfnamefont{H.}~\bibnamefont{Toki}},
  \bibinfo{journal}{Nucl. Phys.} \textbf{\bibinfo{volume}{B435}},
  \bibinfo{pages}{207} (\bibinfo{year}{1995}), \eprint{hep-ph/9312350};
\bibinfo{author}{\bibfnamefont{S.}~\bibnamefont{Sasaki}},
  \bibinfo{author}{\bibfnamefont{H.}~\bibnamefont{Suganuma}}, \bibnamefont{and}
  \bibinfo{author}{\bibfnamefont{H.}~\bibnamefont{Toki}},
  \bibinfo{journal}{Prog. Theor. Phys.} \textbf{\bibinfo{volume}{94}},
  \bibinfo{pages}{373} (\bibinfo{year}{1995}).

\bibitem{tHooft-1981}
\bibinfo{author}{\bibfnamefont{G.}~\bibnamefont{'t~Hooft}},
  \bibinfo{journal}{Nucl. Phys.} \textbf{\bibinfo{volume}{B190}},
  \bibinfo{pages}{455} (\bibinfo{year}{1981}).

\bibitem{lee-1993}
\bibinfo{author}{\bibfnamefont{K.}~\bibnamefont{Lee}}, \bibinfo{journal}{Phys.
  Rev.} \textbf{\bibinfo{volume}{D48}}, \bibinfo{pages}{2493}
  (\bibinfo{year}{1993}).

\bibitem{orland-1994}
\bibinfo{author}{\bibfnamefont{P.}~\bibnamefont{Orland}},
  \bibinfo{journal}{Nucl. Phys.} \textbf{\bibinfo{volume}{B428}},
  \bibinfo{pages}{221} (\bibinfo{year}{1994}), \eprint{hep-th/9404140}.

\bibitem{sato-1995}
\bibinfo{author}{\bibfnamefont{M.}~\bibnamefont{Sato}} \bibnamefont{and}
  \bibinfo{author}{\bibfnamefont{S.}~\bibnamefont{Yahikozawa}},
  \bibinfo{journal}{Nucl. Phys.} \textbf{\bibinfo{volume}{B436}},
  \bibinfo{pages}{100} (\bibinfo{year}{1995}).

\bibitem{akhmedov-1995}
\bibinfo{author}{\bibfnamefont{E.T.}~\bibnamefont{Akhmedov}} \bibnamefont{and}
  \bibinfo{author}{\bibfnamefont{M.A.}~\bibnamefont{Zubkov}},
  \bibinfo{journal}{JETP Lett.} \textbf{\bibinfo{volume}{61}},
  \bibinfo{pages}{351} (\bibinfo{year}{1995});
\bibinfo{author}{\bibfnamefont{E.T.}~\bibnamefont{Akhmedov}},
  \bibinfo{author}{\bibfnamefont{M.N.}~\bibnamefont{Chernodub}},
  \bibinfo{author}{\bibfnamefont{M.I.}~\bibnamefont{Polikarpov}},
  \bibnamefont{and} \bibinfo{author}{\bibfnamefont{M.A.}~\bibnamefont{Zubkov}},
  \bibinfo{journal}{Phys. Rev.} \textbf{\bibinfo{volume}{D53}},
  \bibinfo{pages}{2087} (\bibinfo{year}{1996}), \eprint{hep-th/9505070}.

\bibitem{polyakov-1986}
\bibinfo{author}{\bibfnamefont{A.M.}~\bibnamefont{Polyakov}},
  \bibinfo{journal}{Nucl. Phys.} \textbf{\bibinfo{volume}{B268}},
  \bibinfo{pages}{406} (\bibinfo{year}{1986}).

\bibitem{kleinert-1986}
\bibinfo{author}{\bibfnamefont{H.}~\bibnamefont{Kleinert}},
  \bibinfo{journal}{Phys. Lett.} \textbf{\bibinfo{volume}{B174}},
  \bibinfo{pages}{335} (\bibinfo{year}{1986});
  \textit{ibid.}
  \textbf{\bibinfo{volume}{B246}},
  \bibinfo{pages}{127} (\bibinfo{year}{1990}).

\bibitem{antonov-1998}
\bibinfo{author}{\bibfnamefont{D.}~\bibnamefont{Antonov}} \bibnamefont{and}
  \bibinfo{author}{\bibfnamefont{D.}~\bibnamefont{Ebert}},
  \bibinfo{journal}{Phys. Lett.} \textbf{\bibinfo{volume}{B444}},
  \bibinfo{pages}{208} (\bibinfo{year}{1998}), \eprint{hep-th/9809018};
\bibinfo{author}{\bibfnamefont{D.}~\bibnamefont{Antonov}} \bibnamefont{and}
  \bibinfo{author}{\bibfnamefont{D.}~\bibnamefont{Ebert}},
  \bibinfo{journal}{Eur. Phys. J.} \textbf{\bibinfo{volume}{C8}},
  \bibinfo{pages}{343} (\bibinfo{year}{1999}), \eprint{hep-th/9806153};
\bibinfo{author}{\bibfnamefont{D.}~\bibnamefont{Antonov}} \bibnamefont{and}
  \bibinfo{author}{\bibfnamefont{D.}~\bibnamefont{Ebert}},
  \bibinfo{journal}{Nucl. Phys. Proc. Suppl.} \textbf{\bibinfo{volume}{86}},
  \bibinfo{pages}{456} (\bibinfo{year}{2000}), \eprint{hep-th/9909156}.

\bibitem{baker-1998}
\bibinfo{author}{\bibfnamefont{M.}~\bibnamefont{Baker}},
  \bibinfo{author}{\bibfnamefont{N.}~\bibnamefont{Brambilla}},
  \bibinfo{author}{\bibfnamefont{H.G.}~\bibnamefont{Dosch}}, \bibnamefont{and}
  \bibinfo{author}{\bibfnamefont{A.}~\bibnamefont{Vairo}},
  \bibinfo{journal}{Phys. Rev.} \textbf{\bibinfo{volume}{D58}},
  \bibinfo{pages}{034010} (\bibinfo{year}{1998}), \eprint{hep-ph/9802273}.

\bibitem{dosch-1987}
\bibinfo{author}{\bibfnamefont{H.G.}~\bibnamefont{Dosch}},
  \bibinfo{journal}{Phys. Lett.} \textbf{\bibinfo{volume}{B190}},
  \bibinfo{pages}{177} (\bibinfo{year}{1987}).

\bibitem{simonov-1987}
\bibinfo{author}{\bibfnamefont{Y.}~\bibnamefont{Simonov}},
  \bibinfo{journal}{Nucl. Phys.} \textbf{\bibinfo{volume}{B307}},
  \bibinfo{pages}{512} (\bibinfo{year}{1988}).

\bibitem{antonov-1996}
\bibinfo{author}{\bibfnamefont{D.}~\bibnamefont{Antonov}},
  \bibinfo{author}{\bibfnamefont{D.}~\bibnamefont{Ebert}}, \bibnamefont{and}
  \bibinfo{author}{\bibfnamefont{Y.A.}~\bibnamefont{Simonov}},
  \bibinfo{journal}{Mod. Phys. Lett.} \textbf{\bibinfo{volume}{A11}},
  \bibinfo{pages}{1905} (\bibinfo{year}{1996}), \eprint{hep-th/9605086}.

\bibitem{ball-1988}
\bibinfo{author}{\bibfnamefont{J.S.}~\bibnamefont{Ball}} \bibnamefont{and}
  \bibinfo{author}{\bibfnamefont{A.}~\bibnamefont{Caticha}},
  \bibinfo{journal}{Phys. Rev.} \textbf{\bibinfo{volume}{D37}},
  \bibinfo{pages}{524} (\bibinfo{year}{1988}).

\bibitem{baker-1991}
\bibinfo{author}{\bibfnamefont{M.}~\bibnamefont{Baker}},
  \bibinfo{author}{\bibfnamefont{J.S.}~\bibnamefont{Ball}}, \bibnamefont{and}
  \bibinfo{author}{\bibfnamefont{F.}~\bibnamefont{Zachariasen}},
  \bibinfo{journal}{Phys. Rept.} \textbf{\bibinfo{volume}{209}},
  \bibinfo{pages}{73} (\bibinfo{year}{1991}).

\bibitem{koma-2000}
\bibinfo{author}{\bibfnamefont{Y.}~\bibnamefont{Koma}} \bibnamefont{and}
  \bibinfo{author}{\bibfnamefont{H.}~\bibnamefont{Toki}},
  \bibinfo{journal}{Phys. Rev.} \textbf{\bibinfo{volume}{D62}},
  \bibinfo{pages}{054027} (\bibinfo{year}{2000}), \eprint{hep-ph/0004177}.

\bibitem{gervais-1975}
\bibinfo{author}{\bibfnamefont{J.L.}~\bibnamefont{Gervais}} \bibnamefont{and}
  \bibinfo{author}{\bibfnamefont{B.}~\bibnamefont{Sakita}},
  \bibinfo{journal}{Nucl. Phys.} \textbf{\bibinfo{volume}{B91}},
  \bibinfo{pages}{301} (\bibinfo{year}{1975}).

\bibitem{luscher-1980}
\bibinfo{author}{\bibfnamefont{M.}~\bibnamefont{L\"uscher}},
  \bibinfo{author}{\bibfnamefont{K.}~\bibnamefont{Symanzik}}, \bibnamefont{and}
  \bibinfo{author}{\bibfnamefont{P.}~\bibnamefont{Weisz}},
  \bibinfo{journal}{Nucl. Phys.} \textbf{\bibinfo{volume}{B173}},
  \bibinfo{pages}{365} (\bibinfo{year}{1980}).

\bibitem{baker-2000}
\bibinfo{author}{\bibfnamefont{M.}~\bibnamefont{Baker}} \bibnamefont{and}
  \bibinfo{author}{\bibfnamefont{R.}~\bibnamefont{Steinke}},
  \bibinfo{journal}{Phys. Lett.} \textbf{\bibinfo{volume}{B474}},
  \bibinfo{pages}{67} (\bibinfo{year}{2000}), \eprint{hep-ph/9905375}.

\bibitem{nakahara-diffform}
\bibinfo{author}{\bibfnamefont{M.}~\bibnamefont{Nakahara}},
  \emph{\bibinfo{title}{{Differential forms}}} (\bibinfo{publisher}{Institute
  of Physics Publishing Bristol and Philadelphia}, \bibinfo{year}{1990}), 
  Graduate student series in physics.

\bibitem{koma-1999}
\bibinfo{author}{\bibfnamefont{Y.}~\bibnamefont{Koma}},
  \bibinfo{author}{\bibfnamefont{H.}~\bibnamefont{Suganuma}}, \bibnamefont{and}
  \bibinfo{author}{\bibfnamefont{H.}~\bibnamefont{Toki}},
  \bibinfo{journal}{Phys. Rev.} \textbf{\bibinfo{volume}{D60}},
  \bibinfo{pages}{074024} (\bibinfo{year}{1999}), \eprint{hep-ph/9902441}.

\bibitem{forster-1974}
\bibinfo{author}{\bibfnamefont{D.}~\bibnamefont{F\"orster}},
  \bibinfo{journal}{Nucl. Phys.} \textbf{\bibinfo{volume}{B81}},
  \bibinfo{pages}{84} (\bibinfo{year}{1974}).

\bibitem{davis-1988}
\bibinfo{author}{\bibfnamefont{R.L.}~\bibnamefont{Davis}} \bibnamefont{and}
  \bibinfo{author}{\bibfnamefont{E.P.S.}~\bibnamefont{Shellard}},
  \bibinfo{journal}{Phys. Lett.} \textbf{\bibinfo{volume}{B214}},
  \bibinfo{pages}{219} (\bibinfo{year}{1988}).

\end{thebibliography}
\end{document}